\documentclass[11pt,]{article}
\usepackage[dvips]{graphicx,lscape}
\usepackage{epsfig, subfigure}
\usepackage{psfrag,amsmath}
\usepackage{color,amsthm,amssymb,url}

\usepackage[]{natbib}

\def \IR{\hbox{{\rm I}\kern-.2em\hbox{{\rm R}}}}
\newcommand{\rT}{\mbox{\tiny{T}}}
\newcommand{\N}{\mbox{N}}
\def\mm#1{\ensuremath{\boldsymbol{#1}}} 

\newcommand{\bmy}{\mbox{\boldmath $y$}}

\newcommand{\bmx}{\mbox{\boldmath $x$}}
\newcommand{\bmQ}{\mbox{\boldmath $Q$}}

\newcommand{\bmw}{\mbox{\boldmath $w$}}

\newcommand{\bmS}{\mbox{\boldmath $S$}}

\newcommand{\by}{\mbox{\boldmath $y$}}

\newcommand{\bz}{\mbox{\boldmath $z$}}
\newcommand{\bx}{\mbox{\boldmath $x$}}

\newcommand{\bmzero}{\mbox{\bf 0}}

\newcommand{\bbeta}{\mbox{\boldmath $\beta$}}

\newcommand{\bw}{\mbox{\boldmath $w$}}

\begin{document}
\begin{center}
\begin{Large}
``Spatial Modeling, with Application to Complex Survey Data''.\\

 Discussion of ``Model-based Geostatistics for Prevalence Mapping in Low-Resource Settings", by Diggle and Giorgi
\end{Large}
\vspace{.2in}

\normalsize
Jon Wakefield$^{1,2}$, Daniel Simpson$^3$ and Jessica Godwin$^1$

\vspace{.2in}

$^1$Department of Statistics, University of Washington

$^2$Department of Biostatistics, University of Washington

$^3$Department of Mathematical Sciences, University of Bath
\end{center}

\section{Introduction}

It is a pleasure to discuss the interesting and wide-ranging article of \cite{diggle:giorgi:16}, henceforth referred to as DG.  Prevalence mapping in low resource settings is an increasingly important endeavor  to guide policy making and to spatially and temporally characterize the burden of disease. We will focus our discussion on consideration of the complex design when analyzing survey data, and on spatial modeling. With respect to the former,  we consider two approaches:~direct use of the weights, and a model-based approach using a spatial model to acknowledge clustering. The first of these is considered in Section \ref{sec:surveys}. With respect to spatial modeling we  describe,  in Section \ref{sec:spde}, the  stochastic partial differential equations (SPDEs, Lindgren et al.~2011\nocite{lindgren:etal:11}) approach to  modeling. Throughout, we use  the integrated nested Laplace approximation (INLA, Rue et al.~2009\nocite{rue:etal:09}) to perform computation.  In general, a spatial target of interest may be associated with a point or an area, and in Section \ref{sec:averages} we describe how inference can be made for area averages,  as well as for probabilities of exceedance of a threshold, both using INLA.  A simulation to present the power of the INLA/SPDE approach is provided in Section \ref{sec:simulation}. We conclude with final remarks in Section \ref{sec:discussion}.

\section{Surveys with a Complex Design}\label{sec:surveys}

In the developing world, it is often the case that  disease indicators are collected via complex survey designs. For example, Demographic Health Surveys (DHS) are nationally-representative household surveys that are carried out extensively in the developing world and  typically use a stratified two- or three-stage cluster design \citep{corsi:etal:12}. 
Hence, the data are available with accompanying weights and a randomization-(or design-)based approach to inference is common \nocite{lohr:10}(for a very readable introduction to the analysis of survey data, including randomization-based inference, see Lohr, 2010). In the model-based approach to the analysis of complex survey data \citep{gelman:07}, one accounts for the sampling scheme by including the design (e.g.,~stratification) variables in a regression model. Unfortunately, it is not uncommon for these variables to be unavailable. An alternative approach \citep{chen:etal:14,mercer:etal:14,mercer:etal:15} takes the (asymptotic) sampling distribution of a weighted estimator, such as the Horvitz-Thompson \citep{horvitz:thompson:52} or H{\'a}jek \citep{hajek:71} estimator, as the likelihood and then smooths across space and time.  In practice, often the survey design is ignored in prevalence mapping \citep{dhs-spatial-09:14,bhatt:etal:15}. 

In DG, data from a number of different surveys are analyzed. The rolling malaria indicator survey (rMIS) has a design in which households are randomly selected with a household sampled with probability proportional to village size \citep{roca:etal:12}. School survey data \citep{stevenson:etal:13} are also analyzed by DG; these data do not arise from a standard design, with an iterative process being used for school selection, to limit the chance of overlapping school catchment areas.

We briefly outline the hierarchical modeling approach described in \cite{mercer:etal:14,mercer:etal:15}.
Let $p_{k}$ be the unknown prevalence associated with area $k$,
and let $\widehat{p}_{k}$ be the design-based (weighted) estimator of this prevalence with associated estimated design-based variance  $\widehat{V}^\star_{k}$, $k=1,\dots,K$. The summaries $\{ \widehat{p}_{k}, \widehat{V}^\star_{k} ,k=1,\dots,K \}$ may be obtained using standard software, for example, in Section \ref{sec:simulation} we use the {\tt survey} package \citep{lumley:04} in {\tt R}. The {\tt svyby} function in this package allows the mean prevalence in a region to be estimated, using all of the data collected in that region. The ``data" are then taken as $y_{k} = \log \left( \frac{\widehat{p}_{k}}{1-\widehat{p}_{k}} \right)$ and the asymptotic variance of $y_k$ is obtained, from $\widehat{V}^\star_{k}$, via the delta method, and is denoted  $\widehat{V}_{k}$. If the weighted prevalence estimates are 0 or 1, a fix is required; for example, empirical Bayes may be used. The first stage of the hierarchy is  taken as 
\begin{equation}\label{eq:norm}
y_{k} | \eta_{k} \sim_{iid} \N \left( \eta_{k}, \widehat{V}_{k}\right),
\end{equation}
and smoothing models over space can then be applied to $\eta_k$ to alleviate instability due to small samples, a standard approach in small area estimation (SAE). The use of a normal likelihood based on the empirical logit was used, in a non-complex survey setting, by \cite{stanton:diggle:13}, though with a constant variance. This model is straightforward to fit in {\tt R}, since we can  use {\tt INLA} with a fixed and known variance at Stage 1 of the hierarchy, as in (\ref{eq:norm}). \cite{mercer:etal:15} demonstrate the use of this model when modeling under-5 mortality in Tanzania and we present an example in Section \ref{sec:simulation}.

\section{Building Appropriate Spatial Models}\label{sec:spde}

We now
consider a continuous space Gaussian random field (GRF) model at spatial location $\bx_i$,  with $i$ indexing points at which responses were measured, $i=1,\dots,N$, so that $N$ is the number of data points. For the moment we keep things general, and assume a linear predictor of the form
$$\eta_i = \beta_0 + \bz(\bx_i)\bbeta + S(\bx_i) +\epsilon_i,$$
where  $\bz(\bx_i)$ are covariates measured at the spatial location $\bx_i$, with associated regression coefficients $\bbeta$, $\epsilon_i \sim_{iid} \N(0,\sigma_\epsilon^2)$ is measurement error (aka the nugget term), and $S(\bx_i)$ represents a spatial GRF. 

Many possibilities are available for the form of the covariance  function of the GRF but \cite{stein:99} (amongst others) makes a strong argument for a  Mat{\'e}rn function:
$$
\mbox{cov}[S(\bmx),S(\bmx^\star)] = \frac{\sigma_s^2}{2^{\upsilon-1}\Gamma ( \upsilon) }(\kappa ||\bmx - \bmx^\star ||)^\upsilon K_\upsilon( \kappa ||\bmx - \bmx^\star ||)
$$
where $K_\upsilon(\cdot)$ is a modified Bessel function of the second kind, $\sigma_s^2>0$ is the marginal variance, $\kappa>0$ a scale parameter and $\upsilon>0$ a smoothness parameter. When $\upsilon +1$ is an integer, in two spatial dimensions the Mat{\'e}rn  fields are Markovian \citep{Rozanov1977}. Even in this latter case, data analysis that uses the covariance function directly is  computationally difficult because of the expensive matrix operations that are required \nocite{rue:knorrheld:05}(Rue and Held 2005, Chapter 2).

For modeling the spatial effect, DG use Higdon's convolution kernel approach  \citep{art473}. In order to control the computational complexity inherent in the classical spatial model, the general GRF is replaced with a finite dimensional (or, in their terminology, low-rank) model:
\begin{equation} \label{eqn:basis}
S_n(\bmx) = \sum_{i=1}^n w_i \phi_i(\bmx),
\end{equation}  where the joint distribution of the weights $\mm{w}=[w_1,\dots,w_n]^{\rT}$ is multivariate Gaussian and the deterministic basis functions $\{\phi_i(\bmx)\}_{i=1}^n$ may depend on some parameters being inferred.  The underlying principle is that these finite dimensional random fields will be reasonable proxies for the true latent spatial surface.  The advantage of the finite dimensional representation is that inference costs grow like $\mathcal{O}(n^2N + n^3)$ which, for sufficiently small $n$, is significantly smaller than the $\mathcal{O}(N^3)$ cost of classical methods.  Furthermore, if the basis functions $\phi_i(s)$ are only non-zero in a small part of the domain, the cost is reduced to $\mathcal{O}(N + n^3)$---or $\mathcal{O}(N + n^{3/2})$ for Markovian models---and the method genuinely grows linearly in the number of basis functions \citep{simpson:etal:12a}.   From this point of view, it is clear that kernel methods \citep{art473}, predictive processes \citep{art444}, fixed rank Kriging \citep{art445}, and the SPDE method \citep{lindgren:etal:11} (and for that matter, classical methods like truncated Karhunen-Lo\'{e}ve expansions) are all different faces of the same underlying concept.  The differences between these methods manifest in the way the basis functions and the weights are chosen.  As one would expect, different choices endow these methods with different sets of advantages and disadvantages \citep{bradley2015comparing}.

A particular point that we want to emphasize is that the choice of the spatial random field model is not an innocuous one and this choice will filter through into estimates of uncertainty (be they constructed in  a Bayesian way or not).  In the case where we are interested in predictions at a single unmeasured location,  a small forest of results exist on the  behavior of spatial point predictions for GRFs under the regime in which the data are very close together (infill asymptotics) or in which the data are collected on an expanding domain  \citep{stein:99,zhang2005towards}.  Unfortunately,  for the types of models and applications that DG consider, point estimation is not  the only summary of interest.   In addition, one is interested in estimates of total risk over an area and in locating areas that  exceed a threshold; we consider such endeavors in  Section \ref{sec:averages}.  

Often, data has both a spatial and a temporal component (such as in the rMIS), and in this  case  the number of potential asymptotic regimes that we can use to justify our spatial or spatio-temporal model are dizzying. Even more challenging, is the idea that  for many models the spatial field is designed to model the ``residual'' effect after the potentially non-linear effects of covariates are taken into account.  To the best of our knowledge, the question of how to select the covariance structure of a GRF for the sorts of geostatistical generalized additive mixed models that are increasingly used in practice is completely unstudied.

In our view, the answer is to look for robustness. A little-appreciated fact is that finite-dimensional Gaussian random fields can be spectacularly robust against misspecification.  Why?  Because the non-robustness in GRFs is driven by the very fine-scale effects, which finite dimensional models necessarily discard. To see this, imagine there is a true value of the underlying spatial field $S^*(\bmx)$, which we can write as 
$$
S^\star(\bmx) = \sum_{i=1}^n w_i^\star \phi_i(\bmx) + v^\perp(\bmx),
$$ 
where $v^\perp(\bmx)$ is orthogonal to $\{\phi_i(\bmx)\}_{i=1}^n$.  If the basis functions are chosen appropriately, all of the information in the data can  go  into estimating the main part of the field, which is modeled by the finite-dimensional GRF, while  no assumptions are made about the ``fine-scale'' effects in $v^\perp(\bmx)$, which are smoothed over. Hence, finite-dimensional GRFs will always get the bulk features right at the expense of the fine-scale ones.   
This is different to methods like covariance tapering \citep{art456}, which correctly resolve the fine-scale features necessary for optimal estimation of the field near already observed data points at the expense of resolving the large-scale features \citep{bolin:lindgren:11}.

 This discussion gives  a lot of insight into how we should choose our basis functions.  In  the analysis of the rMIS data in DG, for example, the features of interest were village-level prevalence, which suggests that the basis functions $\phi_i(\bmx)$ should be designed to model features on a village scale. We note that this is slightly different from the  suggestion of increasing the set of basis functions until  inference stabilizes.  We are instead suggesting one looks at the basis functions themselves to see if they can resolve the types of questions that are of interest.  This will lead to very similar results, but is computationally much easier!
 
For point-referenced data, our preferred modeling strategy is the SPDE approach to spatial modeling, as originally described by \cite{lindgren:etal:11}, and subsequently elaborated upon in  \cite{simpson:etal:12a,simpson:etal:12b}.  Rather than choosing basis functions according to the convolution square root of the covariance function, as DG do, we instead focus on classes of functions with good approximation properties.

We now give a brief description of the approach  introduced by \cite{lindgren:etal:11} to approximate Mat{\'e}rn  Markovian Gaussian random fields (MGRFs). The idea is to set up a fine triangular mesh, with $m$ vertices, over the study area. A set of $m$ piecewise linear basis functions $\phi_i(\bmx)$ is then constructed, taking the value 1 at  vertex $i$ and 0 at all other vertices, $i=1,\dots,m$. This gives a set of pyramids that are the building blocks for the approximation. A key point is that these pyramids are non-zero at only a small number of points. The MGRF is again represented by (\ref{eqn:basis}),
with random Gaussian weights $\bw=[w_1,\dots,w_n]^{\rT}$. The spatial prior under this model is therefore, in practice, over functions that are linear combinations of the pyramids (i.e.,~piecewise linear functions over the mesh).  
The flexibility in choosing the triangular mesh allows careful control of how well the spatial effect is resolved.  The general idea is that features that  are more than two triangles large are resolved very well, while those that are smaller than the triangle (such as the value of the field at a point) have a bias of the same order as the triangle size.  For very precise versions of these results, we refer the interested reader to the technical appendices of \citet{simpson2015going}.  

The distribution of $\bw$ is still required, and is chosen to provide a good approximation to the MGRF.
The primary difference between the SPDE approach and the fixed-rank Kriging approach of \citet{art445}, which also recommends using local basis functions chosen for their approximation properties, is the number of parameters that are allowed.  While \citet{art445} aim for a fully flexible model specified with $n(n-1)/2$ parameters, the SPDE approach  focuses instead on a more parsimonious specification with, in the simplest case, only 2 parameters: the scale and the range.  There are obviously computational advantages to this choice, as well as the parsimony of allowing more straightforward specification of meaningful prior distributions \citep{fuglstad2015interpretable}.  The disadvantage is that the two-parameter model, which essentially corresponds to the assumption that the underlying model is isotropic,  is that it may not be flexible enough to correctly model the residual spatial effect.
 
The MGRF that is to be approximated arises as the solution to the SPDE 
\begin{equation}\label{eq:spde}
(\kappa - \Delta)^{\alpha/2} S(\bmx) = \sigma_s W(\bmx),
\end{equation}
where $\Delta = \frac{\partial}{\partial x_1^2} + \frac{\partial}{\partial x_2^2}$ is the Laplacian and $W(s)$ is white noise. The solution to (\ref{eq:spde}) corresponds to a stationary GRF and if $\alpha$ is an integer the GRF is Markovian  \citep{whittle:54}, which is the key for implementation. Note that $\upsilon=\alpha-1$ in the case of a  two-dimensional field and $\alpha=2$ is the default in INLA. In INLA the parameterization is $\theta_1=\log \tau$, $\theta_2=\log \kappa$ where 
$$\tau^2 = \frac{\Gamma(\upsilon)}{\Gamma(\alpha) (4\pi) \kappa^{2\upsilon} \sigma_s^2}.$$

A solution to the SPDE satisfies, for any suitable function $\psi(\bmx)$,
$$\int \psi(\bmx) (\kappa - \Delta)^{\alpha/2} S(\bmx) d\bmx = \sigma_s \int \psi(\bmx) W(\bmx) d\bmx,$$
with these functions are taken to be $\phi_i(\bmx)$, $i=1,\dots,m$. The use of these test functions leads to a system of linear equations to solve, and the solution produces the distribution of $\bmw$ which with a little modification, is a Gaussian Markov random field (GMRF). For the missing details see \cite{simpson:etal:12b}. The GMRF that we obtain for the distribution of the weights comes from two places: the fact that the Mat{\'e}rn form that is used is Markovian and the fact that the basis functions are only non-zero across a small portion of the space.

This prior is combined with the likelihood, with the spatial contribution being  evaluated as a piecewise linear function of the MGRF at the data locations. 
Combining the data $\bmy$ with the above prior gives a  posterior on $\bmS$ which is again of the form (\ref{eqn:basis}),
but with the posterior distribution of the ``weights" being $\bmw | \bmy \sim \N(\bmzero,\bmQ_{S|y}^{-1})$.  In practice, the evaluation of the likelihood at a particular data location turns out to be a weighted sum of the values of the GMRF on the nearest three vertices.  
The above strategy  can be used with a wide range of likelihoods and the SPDE can also  be extended to a variety of non-stationary models \citep{lindgren:etal:11,fuglstad2015does}.

\section{Area Averages and Excursions} \label{sec:averages}

One of the really enjoyable features of DG's paper is their use of continuously specified Gaussian random fields  even when the quantities of interest are areal averages.  We broadly think this is a good idea.  One concrete reason is that integrating  risk over areas allows one to avoid ecological bias, if covariate information is available within areas \citep{wakefield:08}. As the authors point out, however, using such fields is a computational challenge. Our favorite engine for overcoming computational challenges is the {\tt R-INLA} package \citep{rue:etal:09, martins2013bayesian,lindgren2013bayesian}.  For  three types of problem considered in the paper---estimating area-level prevalence, computing areas where prevalence is above a prescribed level, and using spatially-varying models of zero-inflation---{\tt R-INLA} can be used to solve two of them (spatially varying models of zero inflation are beyond the functionality of {\tt R-INLA} for fundamental software design reasons). 

When DG say that  INLA does not provide the joint predictive distribution for the latent field, they are both right and wrong.  By default, INLA computes the univariate predictive distributions and in some cases this is sufficient. It also produces posterior distributions for linear combinations of the latent field, which means that the distribution of the average of the logit prevalence can be obtained. This is, unfortunately, not enough  to compute the joint posterior distribution for area-level prevalences.   Thankfully, the \texttt{R-INLA} package provides a mechanism for sampling from an approximation to the joint posterior distribution, which allows one to estimate the distribution of any functional of the latent field. 
The sampler works by noting that the posterior for the latent Gaussian component, which we will denote by $\mm{\eta}$, can be approximated by   
\begin{equation}\label{posterior.sample}
\tilde{\pi}(\mm{\eta} \mid \mm{y} ) = \int_{\Theta} \pi_G(\mm{\eta} | \mm{y}, \mm{\theta}) \tilde{\pi}(\mm{\theta} | \mm{y}) ~d\mm{\theta},
\end{equation} 
where the hyperparameters are denoted $\mm{\theta}$. In (\ref{posterior.sample}), $ \pi_G(\mm{\eta} \mid \mm{y}, \mm{\theta})$ is the Laplace approximation to the full conditional and $ \tilde{\pi}(\mm{\theta} | \mm{y})$ is the INLA approximation to the posterior of the hyperparameters \citep{rue:etal:09}.  Although this is an, ``integrated Laplace approximation'', the full INLA method proceeds from by using another Laplace approximation to approximate the marginal distributions $\pi({\eta}_j \mid \mm{y} )$. 
The simplest version of the INLA algorithm first constructs a Gaussian approximation to $\pi(\mm{\eta} | \mm{y},\mm{\theta})$ (alternatives includes simplified Laplace and Laplace approximations). This is then used to construct an approximation to $\pi(\mm{\theta} | \mm{y})$, before calculating the approximation $$\pi(\mm{\eta} | \mm{y}) \approx \sum_{i=1}^k w_i\tilde{\pi}(\mm{\eta} | \mm{y},\mm{\theta}_i)\tilde{\pi}(\mm{\theta}_i | \mm{y}),
$$ where $\{(w_i,\mm{\theta}_i)\}_{i=1}^k$ are the weights and points of an integration scheme and $\tilde{\pi}$  stands for the appropriate approximation, that is computed in previous steps of the algorithm \citep{rue:etal:09}.  The \texttt{inla.posterior.sample} functions compute a sample from this approximation to $\pi(\mm{\eta} | \mm{y},\mm{\theta})$ when $\tilde{\pi}(\mm{\eta} | \mm{y},\mm{\theta}_i)$ is the Gaussian distribution that matches the value and curvature of the true conditional distribution at its mode.  To summarize, first the hyperparameters are sampled
from the integration points, $\{\mm{\theta}_i,i=1,\dots,k\}$, and then for each sampled hyperparameter a sample is taken form the Gaussian approximation to the latent field. The alert reader will note that this will not lead to the exact INLA approximation for the marginal distribution $\pi(\eta_j | \mm{y})$, which uses a further Laplace approximation to approximate the univariate conditional distribution $\pi(\eta_i \mid \mm{y},\mm{\theta})$ before integrating out the parameter uncertainty. As these corrected approximations are readily available, they are used to correct the joint mean.  A different option, that we have not yet explored, is to build a copula based around the multivariate approximation that matches the INLA marginals. Our experience is that this sampling algorithm is suitable as it stands, but should the user wish, they could use the returned log-density to compute an independence MCMC sampler that will be asymptotically exact.

The second inferential target that DG consider are excursion sets $\mathcal{A}^+_u = \{ \bmx : p(\bmx) > u\}$, where $p(\bmx)$ is the probability of a binary event of interest and $u$ is some fixed threshold.  Excursion sets are subtle and difficult beasts that have been studied extensively in both the probability \citep{adler:81, book104} and statistics \citep{bolin:lindgren:15,french2016credible} literatures.  The reason these objects are so hard to study is straightforward:~regardless of the statistical philosophy that is being used, the function $p(\bmx)$ is random and hence the set $\mathcal{A}^+_u$ is a random variable.   We would therefore want to find a (say) quantile of $\mathcal{A}^+_u$, for example, the set $A_u^+$ such that $$ \Pr(~p(\bmx)>u\mbox{ for all } \bmx \in A_u^+~) > 1-\alpha,$$ for some prescribed level $\alpha$.  DG's approach to estimation is to  construct the set $\tilde{A}_u^+ = \{\bmx: \Pr(p(\bmx) > u) > 1-\alpha\}$. Since the set is constructed pointwise, this is clearly a multiple testing problem and, in general, the set $\tilde{A}^+_u$ will be too big.  This is because the pointwise tests do not take into account the fact that $p(\bmx)$ is a continuous function and hence there is strong dependence between nearby tests (points): in order for a function value to be above a threshold with high probability (which is the usual case), all of the surrounding points also need to be above that threshold with high probability. This is similar to the reason that care must be taken when simultaneous bands are calculated for unknown functions using splines see, for example,  Wakefield (2013, Section 11.2.7).\nocite{wakefield:13:BOOK}

The situation is even more challenging in the cases that DG have considered due to the complicated sampling design, i.e.,~where the data are spatially located.  In order to say with high probability that a point is above a given threshold, there needs to be a sufficiently large number of observations nearby to narrow down the pointwise uncertainty.  Hence, when you have an inhomogeneous sampling design an excursion set isn't really enough to convey the full information about whether or not you are above a specific threshold.  It would be more useful  to divide the study area into three distinct regions: the upward excursion set $A_u^+$; the downward excursion set $A^-_u$, which  is the set of all points such that $p(\bmx)<u$ with high probability; and the set of points that are in neither the upward nor the downward sets.  This then acknowledges that under imperfect information, there are some areas of the space that you cannot with any certainty say are above or below the threshold.  This type of target cannot be directly computed in \texttt{R-INLA}.  Fortunately, David Bolin has written the excellent \texttt{excursions} package for {\tt R} \citep{bolin:lindgren:15}, which contains a function for computing these regions using output from the \texttt{R-INLA} package. In the next section we illustrate the calculation of both area averages and exceedence probabilities.

\section{Simulation}\label{sec:simulation}

We now demonstrate the power of the SPDE approach as implemented within INLA.  We simulate data within the geography of Kenya, using spatial locations for sampling that correspond to 400 points (enumeration areas) in the 2003 DHS.

For the simulation, we mimic some aspects of the DHS design with enumeration areas (EAs) assumed to be sampled (as first stage clusters) and then households (as second second stage clusters) sampled within EAs. We let $i=1,\dots,n=400$ index the first stage clusters (the EAs), and $j=1,\dots,m_i$, represent households sampled within clusters so that $m_i$ is the number of households in first stage cluster $i$. Let $N_{ij}$ represent the number of household members producing responses in EA $i$ and household $j$, and $Y_{ij}$ be the number of positive responses, $i=1,\dots,n=400$, $j=1,\dots,m_i$. The sampling model is
$$Y_{ij} | p_{ij} \sim \mbox{Binomial}( N_{ij},p_{ij} ),$$
with
$$\mbox{logit }p_{ij} = \beta_0 + S_i + 
\epsilon_{ij},$$
where  $\beta_0$ is the intercept (which relates to the overall log odds of prevalence), $S_i=S_i(\bx_i)$ arises from a spatial model (which we take as an MGRF with variance and range parameters $\sigma_s^2$ and $\kappa$, respectively) and $\epsilon_{ij} \sim_{iid} \N(0,\sigma_\epsilon^2)$ is a random effect that induces dependence between individuals in the same household, in cluster $i$.  In the results we show below, we emphasize that we do not display/include the $\epsilon_{ij}$ terms in prevalence surfaces, as these are assumed to be household specific ``noise". 

The prevalences were generated from a GRF  with mean prevalence of 7\%, so that $\beta_0=\log(0.07/0.93)$. This mean prevalence was chosen based on the national prevalence of HIV in Kenya estimated in the Kenya DHS 2003. The other parameters of the GRF were taken as $\tau =\mbox{e}^{-1/2}$ and $\kappa = \mbox{e}^{1/2}$  with noise variance $\sigma_\epsilon^2=0.01$, in order to produce a prevalence field that approximately matched empirical HIV prevalence estimates from the Kenya DHS 2003 AIDS recode. The practical range, is sometimes defined as the distance at which the correlation drops to 0.13, and is given by $\sqrt{8\upsilon}/\kappa$ \citep{lindgren:rue:15}, which equals 1.72 units here. The marginal variance is $\sigma^2_s=1/(4\pi) = 0.080$.
The number of households in first stage clusters, $m_i$ were taken from the set $(4,5,\dots ,11$) which is a truncated version of the range in the Kenya 2013 DHS.
Denominators (household sizes) $N_{ij}$ were sampled from a discrete distribution on $(1,2,\dots,12)$ also determined by the empirical distribution of the number of people tested per household in the Kenya DHS 2003 AIDS recode. For simplicity, we assume that there are 100 households in each EA.

We let $\pi_i$ be the probability that cluster $i$ is selected, and $\pi_{j|i}$ be the probability that household $j$ is selected, given PSU $i$ was selected, with $i=1,\dots,I$ and $j=1,\dots,m_i$ so that $I$ is the  total number of PSUs (which we take as 46,034, see the methods section of Linard et al.~2010\nocite{linard2010high}) and $m_i$ is the number of SSUs in PSU $i$. The design weight for all individuals within household $j$ of cluster $i$ are taken as the reciprocal of the two-stage cluster sample selection probabilities which are 
$$
\pi_{ij} = \pi_i \times \pi_{j|i} = \frac{400}{46034} \times  \frac{m_i}{100},
$$
where $m_i$ is assumed to be the pre-chosen number of households to select from the 100 households in cluster $i$.

As an illustration of the calculation of area averages, we make inference for the prevalance at the level of ADM1 in Kenya, whose areas  we index by $k$, $k=1,\dots,47$. 
For comparison, and to link with Section \ref{sec:surveys}, in addition to the SPDE GRF, we also fit the hierarchical model in which the first stage is based on the design-based estimate of logit $p_k$, with an associated variance,  for ADM1 area $k$.
Letting $y_{k}$ represent the  logit of the weighted (H{\'a}jek) prevalence, we have $y_k | \eta_k \sim_{iid} \N(\eta_k,\widehat{V}_k)$, and
\begin{equation}\label{eq:ICAR1} \eta_k = \beta_0^\star + S_k+\epsilon_k ,
\end{equation}
where $ \beta_0^\star $ is the area-level intercept, $\epsilon_k \sim_{iid} \N(0,\sigma_\epsilon^2$) are unstructured random effects and $S_k$ are intrinsic conditional autoregressive (ICAR) random effects with variance $\sigma_s^2$. Hence, we are using the popular Besag, York, Molli{\'e} (BYM) model \cite{besag:etal:91}

\begin{figure}\centering
\includegraphics[width=5in,height=3.5in]{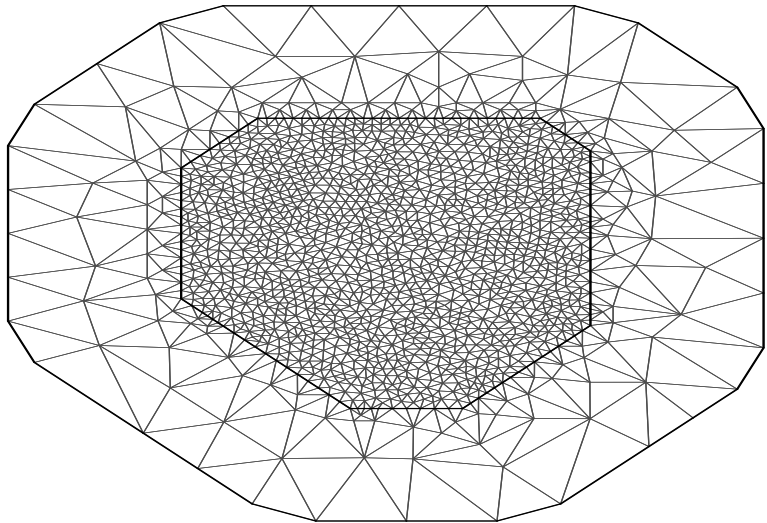}
\caption{ The mesh has two main features.  The first is an inner section, in which the triangles are relatively fine.  This is the area that we are most interested in.  Outside of this inner area, the triangle rapidly become much larger as they get further from the area of interest.  This structure mostly eliminates the boundary effects naturally associated with Markovian models.  \label{fig:mesh}} 
\end{figure}

\begin{figure}
\centering
\includegraphics[width=5.5in,height=4in]{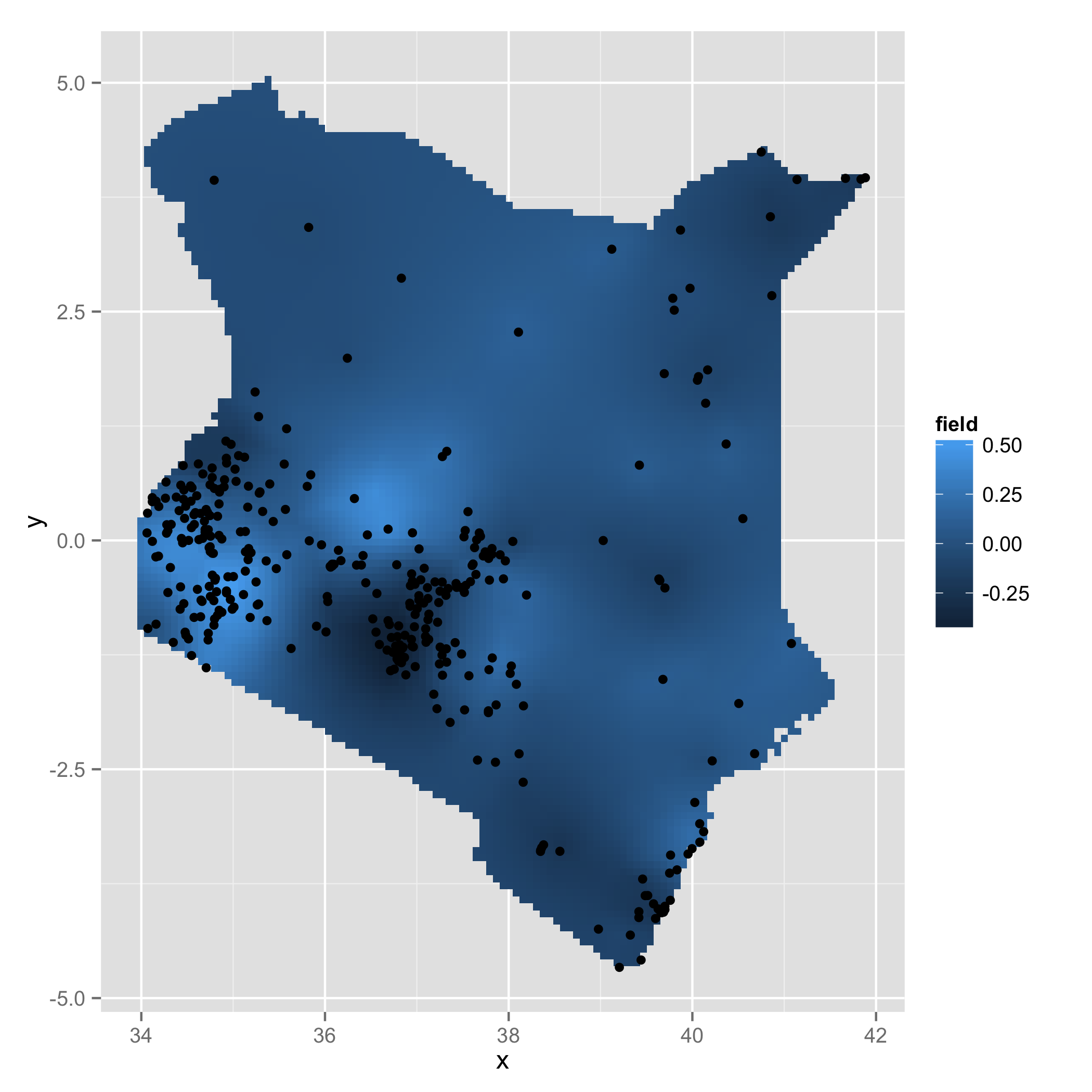}
\caption{ The median posterior spatial effect $S(\bx)$. The smoothness of the field reflects the relatively low amount of information in the data.  In this simulation, points more than $\sqrt{8}/\kappa \approx 1.72$ units apart are essentially uncorrelated.  Hence, in a large part of the space, the estimated spatial contribution is approximately equal to the prior median $0$. At these points, the estimated prevalence  is driven entirely by the country-level mean. \label{fig:median}} 
\end{figure}

\begin{figure}
\centering
\subfigure[True area average prevalence]{
\includegraphics[height=2.8in,width=3.0in]{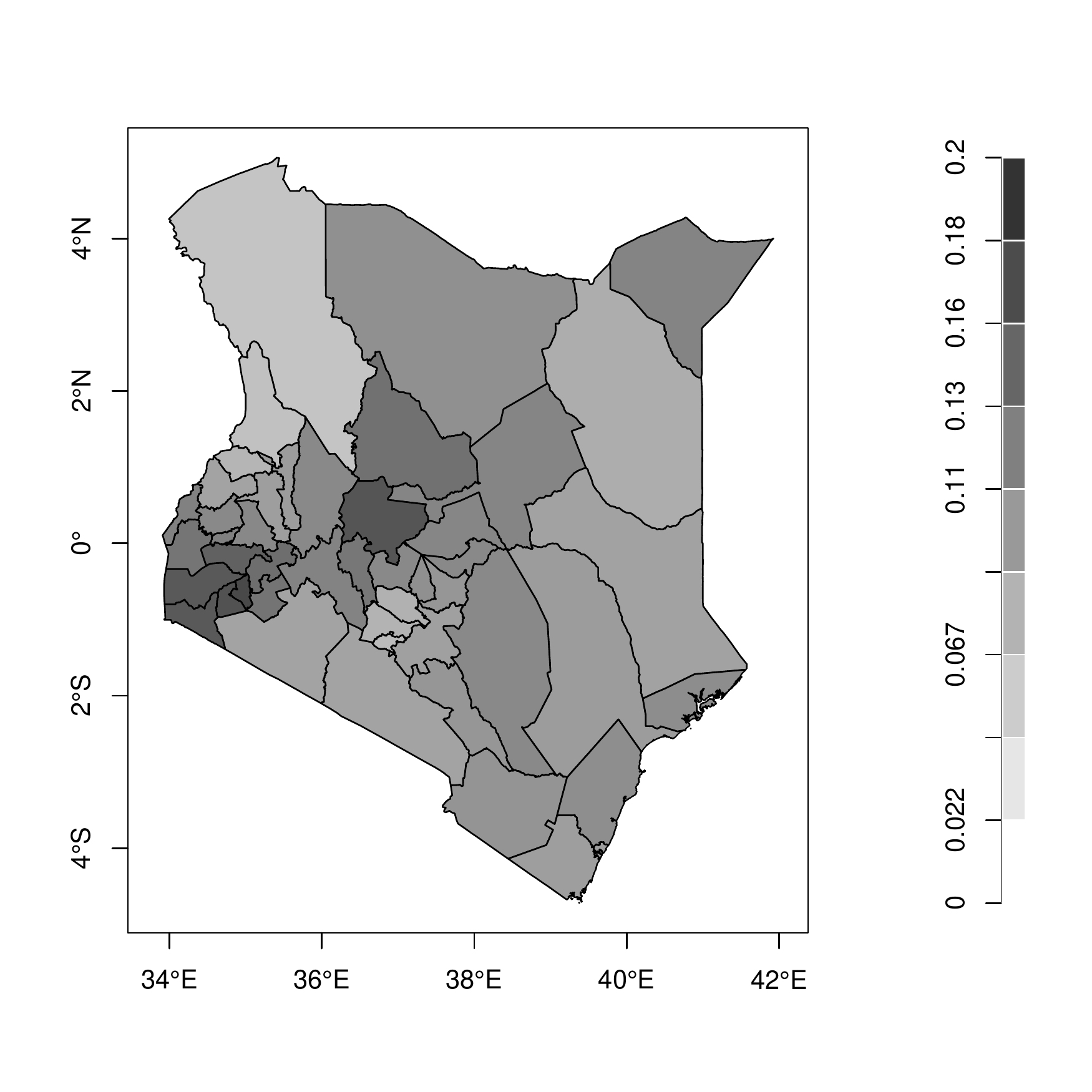}}
\subfigure[Monte Carlo estimate of average, SPDE]{
\includegraphics[height=2.8in,width=3.0in]{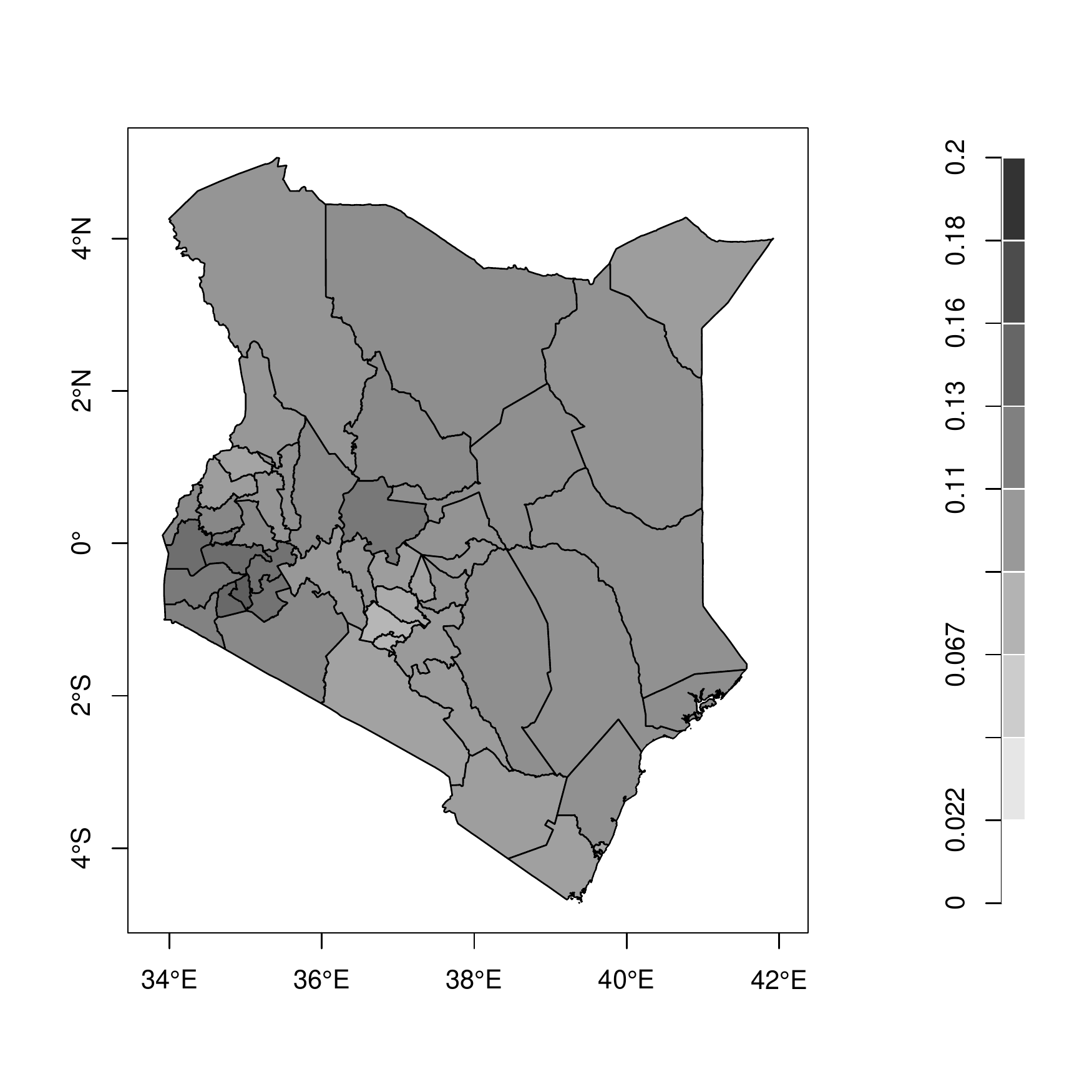}}
\subfigure[Design-based estimate, ICAR]{
\includegraphics[height=2.8in,width=3.0in]{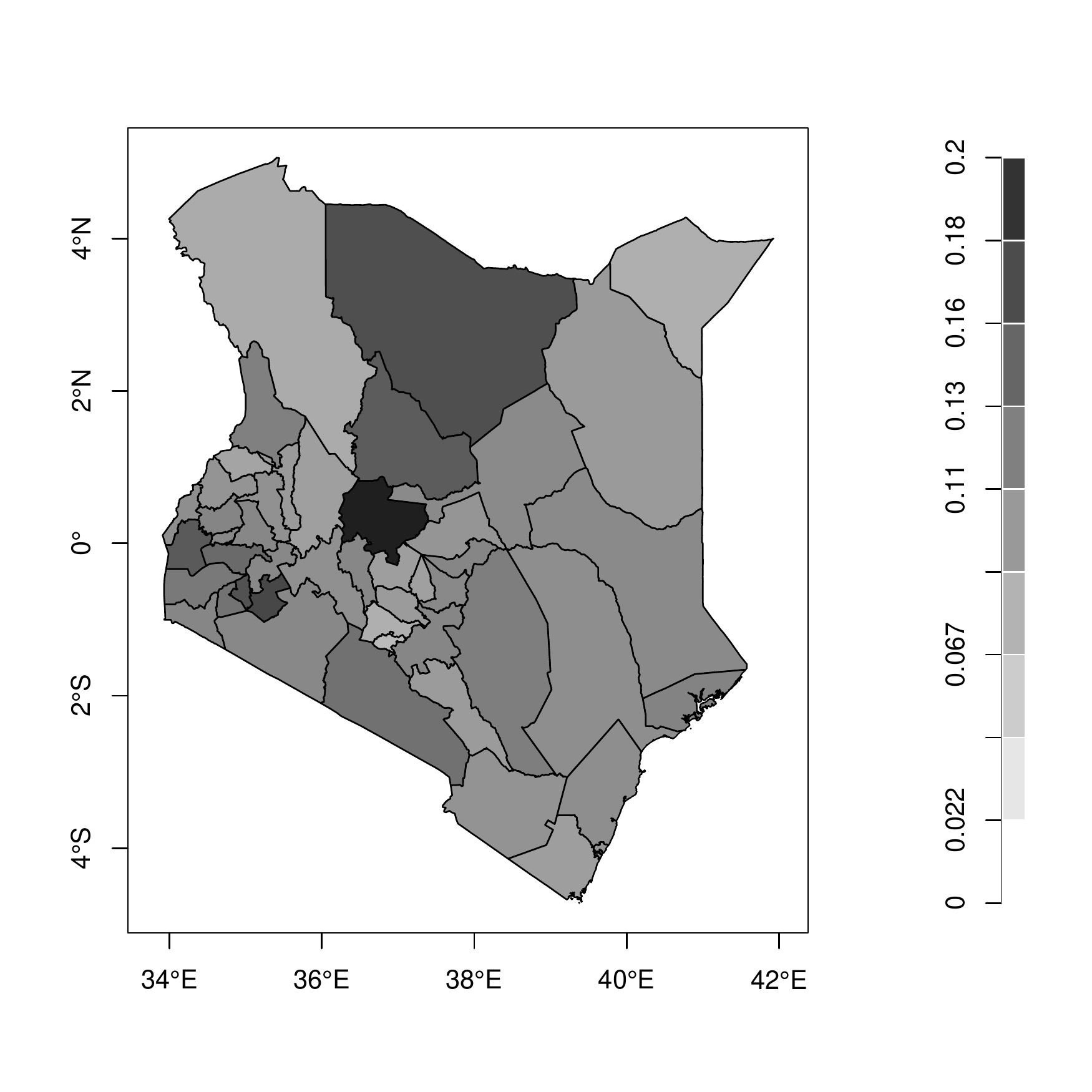}}
\caption{(a) True area-level prevalence averages, (b) estimated area-level prevalences averages from SPDE , (c) estimated area-level prevalences averages from smoothed design BYM model. \label{fig:areas}} 
\end{figure}

The SPDE mesh is shown in Figure \ref{fig:mesh} and in Figure \ref{fig:median} we display the posterior median of  $S(\bx) | \by$, along with the locations at which samples were obtained.   We calculate area-wide summaries of the area level averages,
$$T_k= \int_{A_k} \frac{\exp
\left[ \beta_0+S(\bx) \right]}{ 1+ \exp
\left[ \beta_0+ S(\bx) \right]} ~d\bx$$ where $A_k$ represent the areas in ADM1, $k=1,\dots,K$.
Posterior means of $T_k$ were constructed in \texttt{R-INLA} using Monte Carlo integration with points $\bx_{kj}$, $j=1,\dots,100$, simulated in area $A_k$. Figure \ref{fig:areas}(a) displays the true values of $T_k$ and panel (b) the posterior mean estimates obtained from the SPDE model using the Monte Carlo calculation.  The posterior mean estimates from the design-based approach (including the independent and ICAR random effects) are displayed in  Figure \ref{fig:areas}(c). Overall, the SPDE and smoothed design-based BYM estimates are quite similar, and display some attenuation as compared to the truth.

\begin{figure}
\centering
\includegraphics[width=6.0in,height=4.5in]{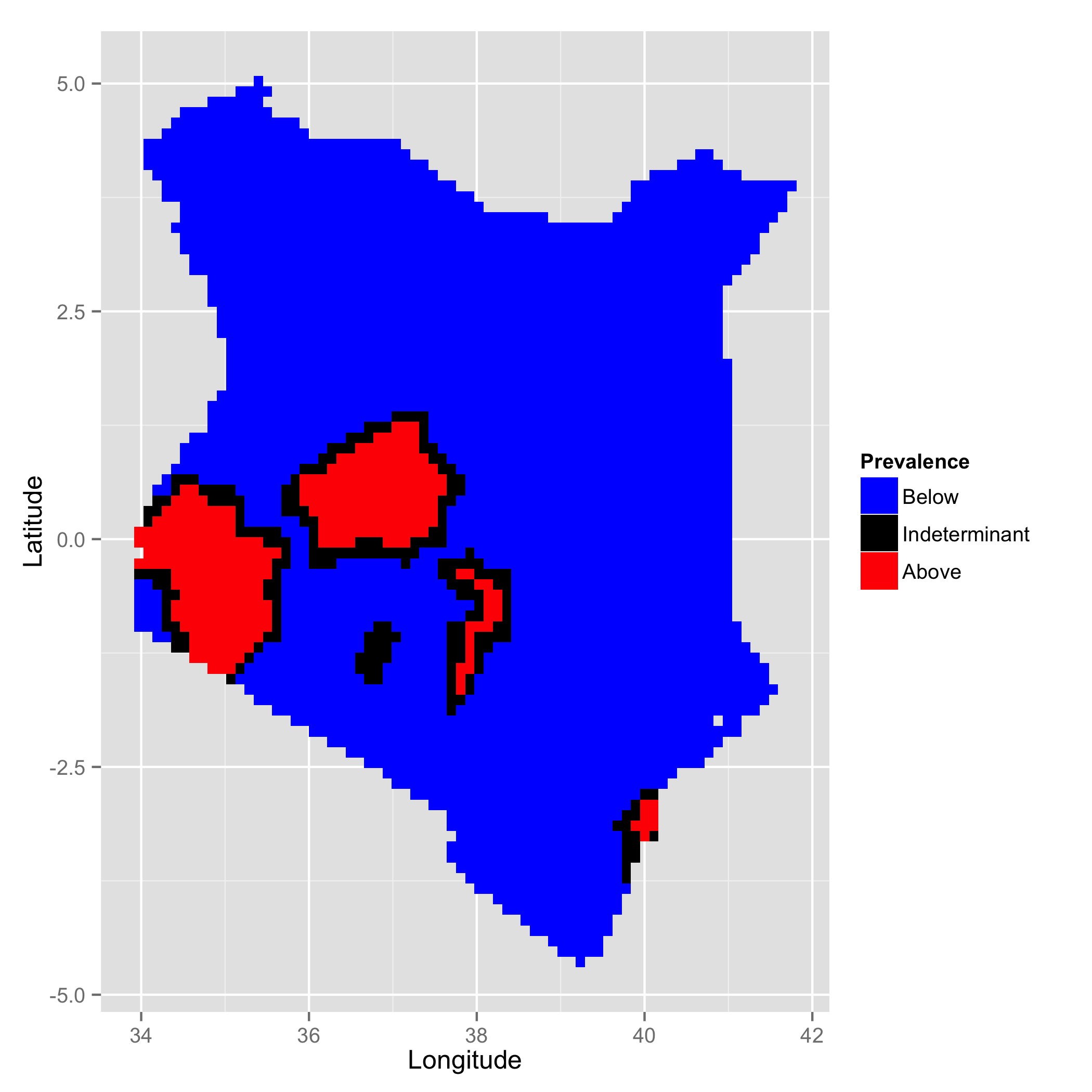}
\caption{The joint posterior regions where the evalence is  simultaneously estimated to be below (blue) or above (red) $7\%$ at a $95\%$ confidence level.  In the black areas, the results are indeterminate. \label{fig:excursion}}
\end{figure}

Figure \ref{fig:excursion} shows the estimate of the $95\%$ excursion sets at the $7\%$ prevalence level calculated using the {\tt excursions.inla} function from the {\tt excursions} package.  The large blue area is such that $\Pr( p(\bmx)<0.07) \text{ for every blue point}) > 0.95$, while the red areas show the  points that simultaneously exceed the threshold.  An interesting features of this figure are the black areas, in which there is not enough information to determine with $95\%$ confidence whether the field is above or below the threshold. 

\section{Concluding Remarks}\label{sec:discussion}

In Section \ref{sec:simulation} we considered a very simple situation in which the design was cluster sampling only. Often, stratification is present also (for example, typically in the DHS there is stratification by urban/rural and perhaps on other variables). In the case of a stratified cluster sampling design, then one may add fixed effects for each of the stratification levels. Post-stratification can also be addressed in the model-based framework  \nocite{gelman:07}(Gelman 2007, Gelman and Hill, 2007, Chapter 14)\nocite{gelman:hill:07}. In addition, covariates can be included, though these need to be known at all locations (at least up to the resolution of the grid) for prediction. Code to reproduce the example in Section \ref{sec:simulation} can be found at \url{http://faculty.washington.edu/jonno/cv.html}.

\bibliographystyle{chicago}
\bibliography{/Users/jonno/Dropbox/BibFiles/spatepi.bib,local.bib}

\end{document}